# A Computational Algorithm for Metrical Classification of Verse

Rama N.[1] and Meenakshi Lakshmanan[2]

[1] Department of Computer Science, Presidency College
Chennai 600 005, India

[2] Department of Computer Science, Meenakshi College for Women
Chennai 600 024, India
and
Research Scholar, Mother Teresa Women's University
Kodaikanal 624 101, India

**Abstract**
The science of versification and analysis of verse in Sanskrit is governed by rules of metre or *chandas*. Such metre-wise classification of verses has numerous uses for scholars and researchers alike, such as in the study of poets and their style of Sanskrit poetical works. This paper presents a comprehensive computational scheme and set of algorithms to identify the metre of verses given as Sanskrit (Unicode) or English E-text (Latin Unicode). The paper also demonstrates the use of euphonic conjunction rules to correct verses in which these conjunctions, which are compulsory in verse, have erroneously not been implemented.

## 1. Introduction

Versification and analysis of poetry in any language have always constituted an exacting combination of art, science and linguistic skill. This is especially so in Sanskrit since it is governed by multitudes of rules to which entire treatises in the ancient literature of the language have been dedicated.

The metrical structure of verses is of prime importance and interest to researchers of Sanskrit literary works. Sanskrit prosody follows a well-defined classification system that identifies verses in terms of its metre (called *chandas* is Sanskrit) and caesura or pauses in verses dictated by sense or natural speech (called *yati* in Sanskrit).

Identification of these two characteristics of a Sanskrit verse serves many purposes for researchers, such as throwing light on the literary acumen of the poet, providing an insight into the metrical preferences of the poet as also creating a knowledge base of the various metres used in a particular treatise. Apart from such and other utilities for the researcher, the identification of the two aforesaid characteristics would also help even a person who is not too knowledgeable in the language and unfamiliar with the meaning of the verse, to chant the verse by pausing at the appropriate junctures.

In this paper, we explore the scheme of classification of Sanskrit prosody [5] and present computational algorithms to classify verses given either in Sanskrit Unicode font or as English E-text.

### 1.1 Unicode Representation

The Unicode (UTF-8) standard is what has been adopted universally for the purpose of encoding Indian language texts into digital format. The Unicode Consortium has assigned the Unicode hexadecimal range 0900 - 097F for Sanskrit characters.

All characters including the diacritical characters used to represent Sanskrit letters in E-texts are found dispersed across the Basic Latin (0000-007F), Latin-1 Supplement (0080-00FF), Latin Extended-A (0100-017F) and Latin Extended Additional (1E00 – 1EFF) Unicode ranges.

The Latin character set has been employed in this paper to represent Sanskrit letters as E-text.

### 1.2 The Basis of the Work

The earliest and most important work in Sanskrit prosody is the *Piṅgala-chandas-śāstra* attributed to the ancient Sage Piṅgala [3]. This treatise consists of *sūtra*-s or aphorisms distributed over eight books [3, 5]. There are various other important and original treatises on the subject that deal extensively with the classification of verse into metres along with the caesura. V. S. Apte's comprehensive work on consolidation of these various sources has been presented by him in [5] and [6], and covers all aspects of Sanskrit prosody. This has therefore been adopted as the primary basis of the present work.

## 2. The Domain

Sanskrit verse is considered a sequence of four *pāda*-s or quarters. Each quarter is regulated either by the number of syllables (*akṣara*-s) or the number of syllabic instants (*mātrā*-s) and the determination of metres is based on either of these factors. Metres based on the first yardstick are called *varṇa* metres, while those based on the second are termed *jāti* metres.

2.1 *Varṇa* Metres

A syllable is as much of a word as can be pronounced at once. There are two types of syllables to contend with – the long (*guru*) and the short (*laghu*). The rule that categorizes letters into these two types is itself given in the form of a verse:

*sānusvāraśca dīrghaśca visargī ca gururbhavet |*
*varṇaḥ saṁyogapūrvaśca tathā pādāntago'pi vā ||*

This succinctly stated rule conveys the following categorization of long and short syllables:

**Short syllables:**

1. Normally, all short vowels – *a, i, u, ṛ, ḷ*.

**Long syllables:**

2. All long vowels – *ā, ī, ū, ṝ*.
3. Any short vowel followed by the *anusvāra* (*ṁ*).
4. Any short vowel followed by the *visarga* (*ḥ*).
5. Any short vowel followed by a double consonant. (The exceptions to this rule are the double consonants *pr*, *br*, *kr* and those starting with *h*. In these four cases, the preceding short vowel can optionally remain short.)
6. Optionally, any short vowel at the end of a *pāda*.

The optional nature of the exceptions mentioned in the rules 5 and 6 above, indicate a sort of poetic license.

From the above discussion it is clear that the four quarters of a verse can each be represented as a sequence of long and short syllables. Traditionally, identification of *varṇa* metres is done on the basis of metrical feet, termed '*gaṇa*-s' in Sanskrit. A *gaṇa* is a combination of three syllables, each of which may be long or short. As such, there are eight such *gaṇa*-s defined as in Table 1, in which 'L' stands for a *laghu* (short) letter, and 'G' for a *guru* (long) one.

Table 1 – *Gaṇa* scheme

| # | Syllable-combination | *Gaṇa* | Corresponding *Gaṇa* in English Poetry |
|---|---|---|---|
| 1 | LGG | *y* | Bacchius |
| 2 | GLG | *r* | Amphimacer |
| 3 | GGL | *t* | Anti-bacchius |
| 4 | GLL | *bh* | Dactylus |
| 5 | LGL | *j* | Amphibrachys |
| 6 | LLG | *s* | Anapaestus |
| 7 | GGG | *m* | Molussus |
| 8 | LLL | *n* | Tribrachys |

In Sanskrit poetry, the number of syllables in a quarter can vary from 1 to 999. When the number of syllables is between 1 and 26 per quarter, the meters are categorized into three:

a. *Sama* (meaning 'equal') – In this, all the four quarters of the verse are identical not just in terms of the number of syllables, but also in the sequence of long and short syllables.

b. *Ardhasama* (meaning 'half-equal') – In this, the first and third quarters are identical, as are the second and fourth.

c. *Viṣama* (meaning 'unequal') – In this, the quarters are uneven or mixed up.

The meters in which there are greater than 26 syllables per quarter are of the '*daṇḍaka*' type and are beyond the scope of this work.

Given that each syllable can be either 'L' or 'G', the number of possible *sama* metres with 1 syllable is $2^1 = 2$, the number with 2 syllables is $2^2 = 4$ and so on. There is clearly a combinatorial explosion in the number of possible 26-syllabled *Sama* metres: $2^{26}$, which is approximately 67 million! Thus, the total number of *sama* metres having anywhere between 1 and 26 syllables per quarter, would be $\sum_{i=1}^{26} 2^i$, which works out to about 134.2 million! For *ardhasama* metres the possible number is obviously even higher, and with *viṣama*, the possibilities are infinite.

However, the number of metres in actual use across the literature is limited to a smaller number than the number theoretically possible. Hence, this work handles the metres in vogue, which indeed themselves constitute a sizeable quantity and pose non-trivial computational problems.

### 2.2 *Jāti* Metres

In this type of metre, each short syllable is counted as constituting one syllabic foot or *mātrā*, while a long syllable is counted as constituting two. Such metres are categorized into two, depending on whether the verse is considered as constituted by two halves, or by four quarters. The various types of the oft-encountered *Āryā* metres are examples of the first variety.

In *Jāti* metres, the identification of metres is done based mainly on the number of certain groups of *mātrā*-s and sometimes partially on *gaṇa* patterns. Standard groups are those of 4, 6, 8 and 10 *mātrā*-s. Groups of 2 and 3 *mātrā*-s are also relevant in a few cases.

It is relevant to state that the same verse could have a *varṇa* classification as well as a *jāti* classification, but this is not necessarily the case.

## 3. The Problem

Given a verse, the problem is to identify its metre and caesura, both as a *varṇa* metre and/or as a *jāti* metre. Once this problem is solved with an efficient algorithm, the same can be applied to classifying the verses of an entire poetic text given as input.

Among the *varṇa* metres, there are close to 600 *sama* metres, 50 *ardhasama* metres and 35 *viṣama* metres in general use in the literature [2]. As for the *jāti* variety of metres, there are about 42 metres in vogue.

The steps in metre classification of a verse are broadly outlined in the algorithm below. This is the way manual identification is done in Sanskrit.

**Algorithm** IdentifyMetre

    Step 1: Parse the verse and identify its four *pāda*-s.

    Step 2: Identify the *guru* (long) and *laghu* (short) syllables.

    Step 3: Identify the *gaṇa* sequence for each *pāda*.

    Step 4: Identify the possible class of *varṇa* metre.

    Step 5: Identify the *varṇa* metre and its caesura based on the *gaṇa*-s, from a database.

    Step 6: Identify the *mātrā* groups and match the pattern with existing *jāti* metres.

end **Algorithm**

The first challenge here has to do with the input verse itself. In Sanskrit, euphonic conjunctions or *sandhi*-s are compulsory in verse. As such, if the input verse does not conform to the *sandhi* rules, the short and long syllables will turn out to be wrongly calculated, and this will result in wrong classification. For example, "*hariḥ iha*" should not be written in this manner, but as "*haririha*", according to the rules of *sandhi* [4]. The *visarga* (*ḥ*) causes the preceding normally short vowel "*i*" to be counted as long. However, when the *sandhi* is properly implemented, the *visarga* disappears, and thereafter the same vowel "*i*" is rightly counted as short. Similar is the case with *sandhi*-s involving the *anusvāra* (*ṁ*). For example, "*phalaṁ ahaṁ*" is a combination of two correctly written words, "*phalaṁ*" and "*ahaṁ*" but when they come together in this way, the combination gets transformed into *phalamaham*, wherein the *anusvāra* is replaced by the ordinary consonant "*m*". The vowel "*a*" preceding the original "*ṁ*" would have been wrongly counted as long if the *sandhi* had not been corrected. Another typical example is *gacchan + iti = gacchanniti*. This sandhi transformation causes the normally short vowel "*a*" occurring just before "*n*", to be counted as long.

One must, however, be careful in the application of *sandhi* rules. In cases such as "*rāmaḥ iti*", which rightly would get transformed into "*rāma iti*", there should not be any further transformation into "*rāmeti*" [4]. At any rate, whether the vowel is "*a*" followed by a *visarga* or is the transformed vowel "*e*", it is counted as long. As such, there is no real problem in this case as far as metre identification is concerned. Hence, such *sandhi* rules need not be applied here.

The second challenge is to identify the four *pāda*-s of the verse. We assume that the input verse has the full stop symbol "|" at the end of the first two *pāda*-s of the verse and a double fullstop symbol "||" at the end of the entire verse, as is the normal practice in Sanskrit poetry. Identifying the two parts of the verse is a trivial task. However, clearly, there are numerous ways of splitting each of these two parts into two *pāda*-s. After all, the metre may be a *sama* metre, an *ardhasama* metre or a *viṣama* metre. Thus, if each half of the verse has 20 syllables, it may be split as 10 syllables per *pāda* or as 9 and 11 or as 8 and 12, etc. It is essential for the solution algorithm to quickly consider the possibilities and arrive at a solution in a reasonable amount of time.

The third challenge lies in the parsing and identification of the short and long syllables according to the rules. This has to be efficiently done. Further, in case no metre is identified with the original *gaṇa* form of the verse, the options that go with the exceptions mentioned in Section 2.1 above need to be considered and tried out.

The fourth challenge is to speed up database lookup, especially given that the keys to be matched are strings.

The fifth challenge is to calculate *mātrā*-s according to the complex rules that govern *jāti* metres.

In addition to this, the sixth challenge is for the classification algorithm to handle both English E-text and Sanskrit Unicode characters as input.

## 4. The Solution

4.1 Handling input given in the Sanskrit Unicode and Latin character sets

As a first step, we convert Sanskrit Unicode character input into English E-text. We then present algorithms to solve the metre classification problem, which accept English E-text [2] as input. The conversion of Sanskrit Unicode characters to Latin characters is based on a Unicode mapping. The complication here is that a single Sanskrit character sometimes maps to a sequence of Latin characters. For example, ᳘ is a single Sanskrit character that is written using the three Latin characters "*vra*". In this work, Unicode mappings were prepared for this purpose of conversion, using the comprehensive Sanskrit 2003 font as basis.

4.2 Algorithm to identify long and short syllables

We now present an algorithm that takes a Sanskrit verse given as input in the Latin Unicode character set, parses it and enumerates the sequence of long and short syllables.

Table 2 – Sets of Sanskrit letters

| # | Name of the Set | Letters in the Set |
|---|---|---|
| 1 | ShortVowels | *a, i, u, ṛ, ḷ* |
| 2 | LongVowels | *ā, ī, ū, ṝ, e, ai, o, au* |
| 3 | Vowels | ShortVowels, LongVowels |
| 4 | Consonants | *k, kh, g, gh, ṅ,*<br>*c, ch, j, jh, ñ,*<br>*ṭ, ṭh, ḍ, ḍh, ṇ,*<br>*t, th, d, dh, n,*<br>*p, ph, b, bh, m* |
| 5 | SemiVowels | *y, r, l, v* |
| 6 | Sibilants | *ś, ṣ, s* |
| 7 | Aspirate | *h* |
| 8 | *Anusvāra* | *ṁ* |
| 9 | *Visarga* | *ḥ* |
| 10 | Columns1and3 (Consonants) | *k, g,*<br>*c, j,*<br>*ṭ, ḍ,*<br>*t, d,*<br>*p, b* |
| 11 | Columns2and4 (Consonants) | *kh, gh,*<br>*ch, jh,*<br>*ṭh, ḍh,*<br>*th, dh,*<br>*ph, bh* |
| 12 | FullStop | \| |

For this, we define that a short syllable has value 0 and a long syllable has value 1. The aim now is to generate the bit sequence for the verse. The following algorithm encapsulates the rules delineated in Section 2.1 above. The categories of Sanskrit letters used in the algorithm have been given in Table 2.

**Algorithm** GenerateBinaryFormOfVerse
// $c_k$ = character in position k in the given text.
// b[ ] is a bit array; j is the next available array index in b.
j = 1;
for k = 1 to LengthOfText
    //Ignore non-vowels and mark long vowels as 1 and
    //short vowels as 0
    if $c_k$ ∈ {Consonants, SemiVowels, Sibilants, Aspirate,
        FullStop, Space, *Anusvāra*, *Visarga*, LineFeed}
       delete $c_k$;
    else if $c_k$ ∈ {LongVowels}
       b[j] = 1; j = j + 1;
    else if $c_k$ ∈ {ShortVowels}
       b[j] = 0; j = j + 1;
    //Check if the short syllable should change to long:
    //1. '*ai*' and '*au*' are long;
    //2. vowel is long if followed by *anusvāra*/*visarga*
    //3. vowel is long if followed by double consonant
    //for special combinations *pr*, *br*, *kr, h* after the
    //vowel, the vowel can be counted as long or short,
    //and hence is marked as an exception.
    if $c_k$ = '*a*' and $c_{k+1}$ ∈ {'*i*', '*u*'}
       b[j] = 1; j = j + 1;
    else if $c_{k+1}$ ∈ {*Anusvāra*, *Visarga*}
       b[j] = 1; j = j + 1;
    else
       if $c_{k+1}$ ∈ {Consonants, SemiVowels, Sibilant,
           Aspirate}
       and $c_{k+2}$ ∈ {Consonants, SemiVowels,
             Sibilants, Aspirate}
          b[j] = 1; j = j + 1;
          if ($c_{k+1}$ ∈ {'*p*', '*b*', '*k*'} and $c_{k+2}$ ∈ {'*r*'}) or
           ($c_{k+1}$ ∈ {Aspirate})
             j is marked as an exception;
          end if
       end if
    end if
end if
//If the double-consonant handled above was from the
//Columns2and4 set, i.e. a Columns1and3 letter
//followed by an Aspirate, then the vowel should be
//reverted to short
if $c_k$ ∈ {ShortVowels} and $c_{k+1}$ ∈ {Columns1and3} and
   $c_{k+2}$ ∈ {Aspirate} and $c_{k+3}$ ∈ {Vowels, FullStop,
   Space, LineFeed}

```
        b[j] = 0; j = j + 1;
    end if
end for
end Algorithm
```

The advantage of this algorithm is that it provides a huge savings in terms of processing, for it eliminates all consonants, etc., which do not play a role in metre determination. As such, words such as *kārtsnyam*, in which there is a concentration of consonants, is processed very quickly. Also, the algorithm proceeds by first eliminating all initial consonant sounds initially, and starts processing only at the first vowel. This again contributes to a savings in computational time. Further, the exceptional cases too are noted for later processing, in a single parse of the verse.

The output produced by the algorithm for the sample *pāda* of a verse, "*vande gurūṇāṁ caraṇāravinde*" is 11011001011.

### 4.3 Identification of *Gaṇa*-s

Reordering and recasting Table 1 in terms of binary values, we have Table 2:

Table 3 – Binary values and decimal equivalents for *Gaṇa*-s

| # | Syllable-combination | Gaṇa | Decimal Equivalent |
|---|---|---|---|
| 1 | 000 | n | 0 |
| 2 | 001 | s | 1 |
| 3 | 010 | j | 2 |
| 4 | 011 | y | 3 |
| 5 | 100 | b | 4 |
| 6 | 101 | r | 5 |
| 7 | 110 | t | 6 |
| 8 | 111 | m | 7 |

The *gaṇa* "bh" has been renamed as "b" just to enhance ease of parsing. Also, in the following discussion, a separate short syllable (value 0) is denoted as "*l*" (for *laghu*) while a long one (value 1) is denoted as "*g*" (for *guru*).

The binary representation that is given as output by the above algorithm is then divided into groups of three, with the last one/two remaining bits kept separately. Thus, the sample output mentioned in Section 4.2 above becomes 110 110 010 11. As per Table 3, the *gaṇa*-s are thus identified as "*ttjgg*", with the last two binary values being retained as single values.

### 4.4 Overall Algorithm to Classify a Verse

Before we identify the *gaṇa*-s as given above, the verse has to be split into four *pāda*-s or quarters. This is done with the following algorithm which utilizes three hash tables, one each for the *Sama, Ardhasama and Viṣama* metres, and two lookup tables, *Ardhasama* Lookup Table (ALT) and *Viṣama* Lookup Table (VLT).

**Algorithm** ClassifyVerse
**Step 1:** Split the verse into two parts ($P_1$, $P_2$) and convert the two parts of the verse into binary form ($B_1$, $B_2$) using Algorithm GenerateBinaryFormOfVerse;
**Step 2:** Generate the set of all possible verse forms $\Lambda$ in binary by permuting the exception information gathered during parsing;
**Step 3:** Let $N_1$ and $N_2$ be the lengths of $B_1$ and $B_2$ respectively. Starting with the binary representation of the original verse form and trying for each of the verse forms, search for the matching *varṇa* meter as follows;
[**Note:**
When split into *pāda*-s, let the binary equivalents of the *pāda*-s be $B_{1,1}$, $B_{1,2}$, $B_{2,1}$, $B_{2,2}$ of lengths $N_{1,1}$, $N_{1,2}$, $N_{2,1}$, $N_{2,2}$ respectively and of *gaṇa* forms $G_{1,1}$, $G_{1,2}$, $G_{2,1}$, $G_{2,2}$ respectively.

The ALT gives the possible ($N_{1,1}$, $N_{1,2}$) values of valid *Ardhasama* metres for a given $N_1$ value. The VLT gives the possible ($N_{1,1}$, $N_{1,2}$, $N_{2,1}$, $N_{2,2}$) values of valid *Viṣama* metres for a given ($N_1$, $N_2$) value.]

```
for each λ ∈ Λ
   [Possibility 1: Equal and even number of syllables in
           P₁ and P₂ – Sama, Ardhasama and
           Viṣama metres are possible]
  if (N₁ = N₂) and (N₁ mod 2 = 0)
    //Sama
    generate pāda-s such that N₁,₁ = N₁,₂ = N₂,₁ = N₂,₂;
    generate set Ψ = {forms of λ | last bit of pāda-s =1};
    for each ψ ∈ Ψ
       generate G₁,₁, G₁,₂, G₂,₁, G₂,₂;
       if B₁,₁ XOR B₁,₂ XOR B₂,₁ XOR B₂,₂ = 0
         if match found on hashing into Sama table
            quit;
         end if
       end if
    end for
    //Ardhasama
    Ω = ALT(N₁);
    if Ω <> ∅ //possible splits exist
      for each ω ∈ Ω
         generate pāda-s B₁,₁, B₁,₂, B₂,₁, B₂,₂ as per ω;
         generate set Ψ={forms of λ | last bit of pāda-s =1};
```

```
         for each ψ ∈ Ψ
            generate G_{1,1}, G_{1,2}, G_{2,1}, G_{2,2};
            if (B_{1,1} XOR B_{2,1} = 0) and (B_{1,2} XOR B_{2,2} = 0)
               if match found on hashing into Ardhasama table
                  quit;
               end if
            end if
         end for
      end for
   end if
   //Viṣama
   Ω = VLT(N_1, N_2);
   if Ω <> ∅ //possible splits exist
      for each ω ∈ Ω
         generate pāda-s B_{1,1}, B_{1,2}, B_{2,1}, B_{2,2} as per ω;
         generate set Ψ={forms of λ | last bit of pāda-s =1};
         for each ψ ∈ Ψ
            generate G_{1,1}, G_{1,2}, G_{2,1}, G_{2,2};
            if match found on hashing into Viṣama table
               quit;
            else if (each of G_{1,1}, G_{1,2}, G_{2,1}, G_{2,2} is a Sama)
               and (match found in list of four special Viṣama)
               quit;
            else if match found in other special Viṣama
               quit;
            end if
         end for
      end for
   end if

[Possibility 2: Equal and odd number of syllables in
               P_1 and P_2 - Ardhasama and Viṣama
               metres are possible]
else if (N_1 = N_2) and (N_1 mod 2 <> 0)
   //Ardhasama
   Ω = ALT(N_1);
   if Ω <> ∅ //possible splits exist
      for each ω ∈ Ω
         generate pāda-s B_{1,1}, B_{1,2}, B_{2,1}, B_{2,2} as per ω;
         generate set Ψ={forms of λ | last bit of pāda-s =1};
         for each ψ ∈ Ψ
            generate G_{1,1}, G_{1,2}, G_{2,1}, G_{2,2};
            if (B_{1,1} XOR B_{2,1} = 0) and (B_{1,2} XOR B_{2,2} = 0)
               if match found on hashing into Ardhasama table
                  quit;
               end if
            end if
         end for
      end for
   end if
   //Viṣama
   Ω = VLT(N_1, N_1);
   if Ω <> ∅ //possible splits exist
      for each ω ∈ Ω
         generate pāda-s B_{1,1}, B_{1,2}, B_{2,1}, B_{2,2} as per ω;
         generate set Ψ={forms of λ | last bit of pāda-s =1};
         for each ψ ∈ Ψ
            generate G_{1,1}, G_{1,2}, G_{2,1}, G_{2,2};
            if match found on hashing into Viṣama table
               quit;
            else if (each of G_{1,1}, G_{1,2}, G_{2,1}, G_{2,2} is a Sama)
               and (match found in list of four special Viṣama)
               quit;
            else if match found in other special Viṣama
               quit;
            end if
         end for
      end if

[Possibility 3: Unequal number of syllables in P_1 and P_2,
               - Viṣama metres possible]
else
   Ω = VLT(N_1, N_2);
   if Ω <> ∅ //possible splits exist
      for each ω ∈ Ω
         generate pāda-s B_{1,1}, B_{1,2}, B_{2,1}, B_{2,2} as per ω;
         generate set Ψ={forms of λ | last bit of pāda-s =1};
         for each ψ ∈ Ψ
            generate G_{1,1}, G_{1,2}, G_{2,1}, G_{2,2};
            if match found on hashing into Viṣama table
               quit;
            else if (each of G_{1,1}, G_{1,2}, G_{2,1}, G_{2,2} is a Sama)
               and (match found in list of four special Viṣama)
               quit;
            else if match found in other special Viṣama
               quit;
            end if
         end for
      end for
   end if
  end if
end for
end Algorithm
```

In the above algorithm, the ALT and VLT save computational time, because there are numerous ways of dividing the $N_1$ number syllables of $P_1$ into two unequal parts. Similar is the case with the $N_2$ number of syllables of $P_2$. Now the ALT and VLT return a set of possible split sets for Ardhasama and Viṣama metres respectively, given $N_1$ and $N_2$. Thus, by only splitting $B_1$ and $B_2$ into pāda-s in these known possible ways, rather than in numerically all possible ways, the solution search space is considerably reduced, thus reducing the computational time. Furthermore, the binary representation of the syllables enables the use of binary arithmetic (such as the use of the operator XOR), which can speed up operations in contrast to letter-wise comparison.

### 4.5 Hashing

The hash table for *Sama* metres is organized as one with 26 buckets, the j[th] bucket holding metres having j syllables per *pāda*. Using the input which is a *gaṇa* sequence such as "ttjgg" in the example cited earlier, we first find the appropriate bucket and then hash into the value [1].

Let each *gaṇa* in the sequence be denoted as $G_i$ and the last one or two bits be denoted as $B_j$. Thus the input to the hashing function will be of the form $G_1G_2G_3\ldots G_nB_1\ldots B_m$ where clearly, n can take a maximum value of 8 and m, a maximum of 2. We now find the decimal equivalent of each $G_i$'s binary value as per Table 3. However, for *gaṇa* sequences like *nnn* and *nn*, which are the binary sequences 000 000 000 and 000 000 respectively, we get the decimal equivalent 0 for both. To therefore avoid hashing collisions, we first append 1 as the most significant bit to each $G_i$'s 3-bit equivalent, making each a 4-bit number. Let the new binary sequence be denoted by $R_1R_2R_3\ldots R_nB_1..B_m$, where each $R_i$, $1<= i <= n$, consists of 4 bits. We then find the decimal equivalent $C_i$ for each $R_i$. Thus we obtain the new sequence of decimal numbers $C_1C_2C_3\ldots C_nB_1\ldots B_m$. The hash key value is calculated as follows:

$$\text{HashKey} = \text{BSA} + \sum_{i=1}^{n} C_i * 16 + \sum_{i=1}^{m} B_i * 2$$

where BSA is the Bucket Starting Address given by

$$\text{BSA} = \sum_{i=1}^{k-22} p_1 + \sum_{i=k-22}^{k-6} p_2 * \frac{3|i|-i}{2|i|} + \sum_{i=k-6}^{k-1} p_1 * \frac{3|i|-i}{2|i|}$$

Here, $p_1 = 53$ and $p_2 = 103$ are empirically determined primes. The scheme uses 16 as a multiplier for the $C_i$'s because the Binary Coded Decimal scheme is followed for the $C_i$'s, and 2 as a multiplier for the $B_i$'s because the $B_i$'s are all single binary digits.

The metre identification is done in O(1) time. With the established values of $p_1$ and $p_2$, there is found to be a maximum number of collisions of three, and that too for very few metres. Collisions were handled using the chaining method. In all, the maximum number of string comparisons required to zero-in on the metre is only 3, and this maximum is actually reached only in very few cases.

The hash table for the *Ardhasama* metres is designed according to the syllables in the first two *pāda*-s while that for the *Viṣama* metres is designed according to the syllables in the four *pāda*-s of the verse. The hash table itself provides the caesura for each metre along with the name of the metre. Thus, the pause-points in each verse are got at once along with the name of the metre.

### 4.6 Algorithm for *Jāti* Metres

The binary sequence earlier computed for the two verse parts are converted into a decimal sequence with 1's converted to 2's and 0's converted to 1's. The algorithm for *Jāti* metres adds successive values to check for the *mātrā* combinations provided for all the *jāti* metres, one by one. Since the rules are complex and include rules that prohibit overlap of *mātrā*-s across letters, this rule-based approach rather than the hash table approach was found more effective.

## 5. Conclusions

The benefits of the presented algorithm are many. One parse of the E-text alone is carried out, all possible exceptional cases are handled, and the possibilities of *sama, ardhasama and viṣama* metres as well as *jāti* metres are handled. Further, the initial scheme for binary conversion ensures that only vowels are processed, and enables the use of binary operators for comparison, thus speeding up the computation considerably. Moreover, the computational schema presented earlier by the authors [4] to form euphonic conjunctions from given words, finds application here - input verses incorrectly given without the euphonic conjunctions having been handled, are corrected by this algorithm before processing begins. Another advantage with the presented schema is that the input E-text can be either as Sanskrit Unicode characters or as Latin Unicode characters.

Thus, the presented computational algorithm provides, for the first time, a solution to the non-trivial computational problem of metre identification in the realm of automated Sanskrit text processing. It uses a novel schema and efficient search methods to speedily and yet comprehensively achieve the aim of identifying the metre and caesura of given Sanskrit verses.

## References

[1] Donald E. Knuth, The Art of Computer Programming Volume 3: Sorting and Searching, Addison Wesley, Second Edition, 1998.
[2] Göttingen Register of Electronic Texts in Indian Languages (GRETIL), www.sub.uni-goettingen.de/ ebene_1 /fiindolo/gretil.htm.
[3] Piṅgala, *Chandas-śāstra*, Kāvyamālā Series No. 91 (3[rd] Edition), Bombay, 1938.
[4] Rama N., Meenakshi Lakshmanan, A New Computational Schema for Euphonic Conjunctions in Sanskrit Processing, IJCSI International Journal of Computer Science Issues, Vol. 5, 2009 (ISSN (Online): 1694-0784, ISSN (Print): 1694-0814), Pages 43-51.
[5] Vaman Shivram Apte, Practical Sanskrit-English Dictionary Appendix A.I – Sanskrit Prosody, Motilal

Banarsidass Publishers Pvt. Ltd., Delhi, 1998, Revised and Enlarged Edition, 2007.

[6] Vaman Shivram Apte, Practical Sanskrit-English Dictionary Appendix A.II – A Classified List of Sanskrit Metres, Motilal Banarsidass Publishers Pvt. Ltd., Delhi, 1998, Revised and Enlarged Edition, 2007.

**Dr. Rama N.,** B.Sc. Mathematics (1986), Master of Computer Applications (1989) and Ph.D. Computer Science (2003). She served in faculty positions at Anna Adarsh College, Chennai, India and as Head, Department of Computer Science at Bharathi Women's College, Chennai, before moving on to Presidency College, Chennai, where she currently serves as Associate Professor. She has 20 years of teaching experience including 10 years of postgraduate (PG) teaching, and has guided 15 M.Phil. students. She has been the Chairperson of the Board of Studies in Computer Science for UG, and Member, Board of Studies in Computer Science for PG and Research at the University of Madras. Current research interests: Program Security. She is the Member of the Editorial cum Advisory Board of the Oriental Journal of Computer Science and Technology.

**Meenakshi Lakshmanan** B.Sc. Mathematics (1996), Master of Computer Applications (1999), M.Phil. Computer Science (2007).Currently pursuing Ph.D. Computer Science at Mother Teresa Women's University, Kodaikanal, India. She has also completed Level 4 Sanskrit (*Samartha*) of the prestigious *Samskṛta Bhāṣā Pracāriṇī Sabhā*, Chittoor, India. Starting off her career as a software engineer at SRA Systems Pvt. Ltd., she switched to academics and currently heads the Department of Computer Science, Meenakshi College for Women, Chennai, India. She is a professional member of the ACM and IEEE.